\documentclass[a4paper,11pt]{article}
\usepackage{jcappub} 
\usepackage[capitalize]{cleveref}
\usepackage{glossaries}
\usepackage{subcaption}
\usepackage{macros} 
\usepackage{bm}
\usepackage{floatrow}
\usepackage{sidecap}

\usepackage[dvipsnames]{xcolor}
\definecolor{blue}{rgb}{0.16, 0.34, 0.76}
\definecolor{amaranth}{rgb}{0.9, 0.17, 0.31}
\definecolor{pink}{rgb}{0.87, 0.56, 0.81}
\definecolor{forestgreen}{rgb}{0.0, 0.5, 0.0}
\definecolor{maroon}{rgb}{0.76, 0.13, 0.28}
\definecolor{cardinal}{rgb}{0.77, 0.12, 0.23}
\definecolor{lightcardinal}{rgb}{0.97, 0.42, 0.53}
\definecolor{frenchlila}{rgb}{0.53, 0.38, 0.56}
\definecolor{yellow}{rgb}{1.0, 1.0, 0.87}
\definecolor{lightseagreen}{rgb}{0.7, 0.92, 0.68}
\definecolor{gray}{rgb}{0.9, 0.9, 0.9}
\definecolor{lightblue}{rgb}{0.66, 0.84, 0.96}

\hypersetup{citecolor=blue,
		colorlinks=true,
		linkcolor=blue,
		urlcolor=blue}

\newcommand{\param}{\ensuremath{\boldsymbol{\theta}}}
\newcommand{\baralpha}{\ensuremath{\overline{\alpha}}}
\newcommand{\ghyp}{\ensuremath{\mathcal{GH}}}
\newcommand{\hyp}{\ensuremath{{\cal H}}}
\newcommand{\stt}{\ensuremath{{\cal T}}}
\newcommand{\whtl}{\ensuremath{{\cal W}}}
\newcommand{\gaus}{\ensuremath{{\cal N}}}

\newcommand{\mrm}[1]{\mathrm{#1}}

\newacronym{lisa}{LISA}{Laser Interferometer Space Antenna}
\newacronym{esa}{ESA}{European Space Agency}
\newacronym{gw}{GW}{gravitational wave}
\newacronym{sgw}{SGW}{stochastic gravitational wave}
\newacronym{sgwb}{SGWB}{stochastic gravitational-wave background}
\newacronym{gr}{GR}{general relativity}
\newacronym{tdi}{TDI}{Time-Delay Interferometry}
\newacronym{mosa}{MOSA}{movable optical sub-assembly}
\newacronym{oms}{OMS}{optical metrology system}
\newacronym[longplural={power spectral densities}]{psd}{PSD}{power spectral density}
\newacronym[longplural={amplitude spectral densities}]{asd}{ASD}{amplitude spectral density}
\newacronym[longplural={cross spectral densities}]{csd}{CSD}{cross spectral density}
\newacronym{rms}{RMS}{root mean square}
\newacronym{snr}{SNR}{signal-to-noise ratio}
\newacronym[longplural={discrete Fourier transforms}]{dft}{DFT}{discrete Fourier transform}
\newacronym{map}{MAP}{maximum a posteriori estimate}
\newacronym{dof}{DoF}{degrees of freedom}
\newacronym{mcmc}{MCMC}{Markov chain Monte Carlo}
\newacronym{rj}{RJ}{reversible jump}
\newacronym{pta}{PTA}{pulsar timing array}
\newacronym{ipta}{IPTA}{International Pulsar Timing Array}
\newacronym{ucbs}{UCBs}{Ultra Compact Galactic Binaries}
\newacronym{pt}{PT}{Phase-transitions}

\arxivnumber{1234.56789} 
\title{Characterization of non-Gaussian stochastic signals with heavier-tailed likelihoods}

\author[a]{N. Karnesis}
\author[a,b]{A. Sasli}
\author[c]{R. Buscicchio}
\author[a]{N. Stergioulas}
\affiliation[a]{Department of Physics, Aristotle University of Thessaloniki,\\
Thessaloniki 54124, Greece}
\affiliation[b]{School of Physics and Astronomy, University of Minnesota,\\
55455 MN, USA}
\affiliation[c]{Dipartimento di Fisica ``G. Occhialini'', Universit\`a degli Studi di Milano-Bicocca, \\
Piazza della Scienza 3, 20126 Milano, Italy}

\emailAdd{karnesis@auth.gr}

\abstract{Future Gravitational Wave observatories will give us the opportunity to search for stochastic signals of astrophysical, or even cosmological origins. However, parameter estimation and search will be challenging, mostly due to the overlap of multiple signal components, as well as the potentially partially unknown properties of the instrumental noise. In this work, we propose a robust statistical framework based on heavier-tailed likelihoods for the characterization of stochastic gravitational-wave signals. In particular, we use the symmetric hyperbolic likelihood, which allows us to probe the signal spectral properties and simultaneously test for any departures from Gaussianity. We demonstrate this methodology with synthetic data from the future LISA mission, where we estimate the potential non-Gaussianities induced by the unresolved Ultra Compact Galactic Binaries.}

\begin{document}
\maketitle
\flushbottom

\section{Introduction
\label{sec:intro}}

Since the dawn of \gls{gw} astronomy in the last decade~\citep{firstGW, GWTC,GWTC1,GWTC-2,GWTC-4,GWTC-3,GWTC-5}, one of our objectives is to directly measure the \gls{sgw} component of a signal. This can have both an astrophysical or a cosmological origin, and it can be present across the frequency spectrum~\cite{Babak:2023lro, Amaro2017, PhysRevD.104.043019, sym14020270, Caprini_2018, LISACosmologyWorkingGroup:2022jok}. The expectations are high given the upcoming \gls{gw} observatories, both using laser interferomentry~\cite{KAGRA,Unnikrishnan,ET,CE,Amaro2017,chinesedetectors}, and pulsar timing arrays~\cite{NANOGrav2020bcs,NANOGrav2021flc,Antoniadis:2022pcn}. At nano-Hz frequencies, the first evidence for an unresolved \gls{gw} signal of astrophysical and/or cosmological origin~\cite{2023ApJ...951L..11A,2024A&A...685A..94E} has recently been reported by the European pulsar timing array~\cite{2023A&A...678A..50E}, NANOGrav~\cite{2023ApJ...951L...8A}, the Parkes Pulsar timing array~\cite{2023ApJ...951L...6R} and the Chinese pulsar timing array~\cite{2023RAA....23g5024X}. However, especially for space-based observatories such as the \gls{lisa}\footnote{The \gls{lisa} mission was ``adopted'' by the \gls{esa} during the writing of this paper, on 25 January 2024~\cite{lisa_adoption}.}, the search and characterization of \gls{sgwb} signals is expected to be challenging. \gls{lisa} is going to be signal dominated~\cite{Amaro2017,LISACosmologyWorkingGroup:2022jok}, which means that an abundance of transient signals, as well as multiple overlapping \gls{sgwb} components are going to populate the data streams. In some cases, the multiple overlapping signals are expected to generate confusion stochastic signal components, which will increase the uncertainty in our knowledge of the instrumental noise, further complicating the analysis. Moreover, for the case of a single detector, cross-correlation analyses between the measured data are not possible~\cite{LIGOScientific:2019vic,Dhurandhar:2007vb,Renzini:2022alw}. At the same time, the presence of noise non-stationarities, such as fast noise transients (glitches)~\cite{Baghi:2021tfd,Powell:2018csz}, might introduce non-Gaussian components. This means that accurate models for the \gls{sgwb}s and instrumental noise statistical properties are going to be crucial to fully understand the data~\cite{Speri:2022kaq}.

There exists extensive literature dedicated to the treatment of non-Gaussian signals. Some of the proposed approaches focus on the detection and subtraction of the quantities that cause departures from Gaussian noise (for example, detecting and subtracting glitches~\cite{Baghi:2021tfd,Powell:2018csz,LIGOScientific:2019hgc,Cornish:2020dwh}), while others follow the route of describing the given data outliers statistically, i.e., modeling them on the likelihood level~\cite{roever2011A,roever2011B,Sasli:2023mxr}. The latter is very useful in various situations, for example, when models for transient noisy signals (glitches) are not available, or when the residuals from subtracting louder sources can not be fully described, or finally, when the stochastic signal itself has non-Gaussian properties~\cite{Buscicchio:2022raf,Bartolo:2018qqn,Ballelli:2022bli}. 

This is the approach that we follow in this work, adapting the methodology based on the heavier-tailed hyperbolic distribution, which was first introduced in the context of \gls{gw} data analysis in~\cite{Sasli:2023mxr}. This framework allows us to fully characterize a given stochastic component of the noise (whether it originates from the instrument and local environment, or the Universe), inferring both its spectral shape, as well as testing for any departures from Gaussianities simultaneously. This data analysis methodology can be applied to an arbitrary number of detectors or data channels, providing information about the statistical properties of the data across the spectrum. 

This work is organized as follows. In Section~\ref{sec:theory}, we introduce the theoretical background of the methodology. Then, in Section~\ref{sec:toys} we demonstrate the proposed framework with simple synthetic data sets. In Section~\ref{sec:lisa} we apply our methodology to a more realistic scenarios, using synthetic \gls{lisa} data. Finally, we summarize and discuss our findings in Section~\ref{sec:discussion}.  

\section{Theoretical Background
\label{sec:theory}} 

\subsection{The Generalized Hyperbolic distribution\label{sec:generalized_hyperbolic_distr}}

We begin with a description of the family of Generalized Hyperbolic (\ghyp) distributions~\cite{Bianchi2020, Borak2011,Prause1997ModellingFD, Eberlein2002TheGH, Prause1999TheGH, Sasli:2023mxr}, which is essentially a parametric family of distributions that has a great selection of the most used exponential distributions as limiting distributions. Thus, one of the interesting attributes of \ghyp{}, is that we can use it to arrive at virtually any distribution of the exponential family, by suitably choosing its parameters. Following the notation of~\cite{Sasli:2023mxr, Prause1999TheGH}, the multivariate \ghyp{} for a variable $\bm{x}\in {\rm I\!R}^d$, with $\bm{\mu}, \bm{\beta} \in {\rm I\!R}^d$ reads:
\begin{align}
{\rm \cal GH}_{d} & (\bm{x} | \lambda, \alpha, \bm{\beta}, \delta, \bm{\mu}) = \nonumber \\ 
& \frac{\left( \sqrt{\alpha^2 - B}/\delta\right)^d}{(2\pi)^{d/2} K_\lambda \left( \delta \sqrt{\alpha^2 - B}\right)} \frac{K_{\lambda - d/2} \left( \alpha \sqrt{\delta^2 + r}\right)}{\left( \alpha^{-1}\sqrt{\delta^2 + r}\right)^{d/2 - \lambda}} \exp \left[ \bm{\beta}^{\rm T} (\bm{x}-\bm{\mu})\right],
\label{eq:genhypmd}
\end{align}
where we have defined $B = \bm{\beta}^{\rm T} \Delta^{-1}\bm{\beta}$, and $r$ represents the residuals 
\begin{equation}
    r = (\bm{x} - \bm{\mu})^{\rm T} \Delta^{-1} (\bm{x} - \bm{\mu}) .
\label{eq:res}
\end{equation}
$K_\lambda$ denotes the modified Bessel function of the third kind of order $\lambda$~\cite{Sasli:2023mxr, Prause1999TheGH}. The parameters of \ghyp{} $(\lambda, \alpha, \bm{\beta}, \delta, \bm{\mu})$ take values
\begin{equation}
\begin{array}{ll}
\mu \in \mathbb{R} \\
\delta \geq 0,|\bm{\beta}|<\alpha, & \text { if } \lambda>0, \\
\delta>0,|\bm{\beta}|<\alpha, & \text { if } \lambda=0, \\
\delta>0,|\bm{\beta}| \leq \alpha, & \text { if } \lambda<0 .
\end{array}
\label{eq:domain}
\end{equation}
The parameter $\mu$ controls the location of the mode of the distribution, while $\beta$ describes its skewness, with $\beta=0$ corresponding to a symmetric distribution. Finally, the parameters $\alpha$ and $\delta$ control the overall shape of the distribution. The term $\Delta$ is a positive definite unitary matrix $\in {\rm I\!R}^{d\times d}$. It encapsulates the correlation between the data points of the measured time-series:  in our particular case, it contains information about the cross-correlation between different data channels. 

As already mentioned, quite a few commonly-used distributions are limiting cases of the \ghyp{} distribution. Some notable examples are the Gaussian (\gaus{}), the Student-t (\stt{}) and the symmetric Hyperbolic (\hyp{}) distributions, which we will focus on the applications of Sections~\ref{sec:toys} and~\ref{sec:lisa}. The \hyp{}, for the uni-variate case is derived by setting $\lambda=1$ and $\beta=0$. The \ghyp{} converges to the Gaussian distribution $\gaus(\mu, \sigma^2)$ when $\lambda=1,\, \beta=0$ and $\delta,\,\alpha \rightarrow \infty$, in such a way that $\delta / \alpha \rightarrow \sigma^2$. Finally, the \stt{} with $\nu$ degrees of freedom is a limiting case of the \ghyp{} distribution when $\lambda=-\nu/2, \alpha= 0, \beta= 0, \delta=\sqrt{\nu}$ and $\mu=0$. More details on the limiting distributions of \ghyp{} can be found in~\cite{ernst, Bianchi2020, Borak2011,Prause1997ModellingFD, Eberlein2002TheGH, Prause1999TheGH, Sasli:2023mxr} and references therein.

Taking into account the above, the flexibility of the \ghyp{} distribution shows great potential for the description of complex data, which, as shown in~\cite{Sasli:2023mxr}, comes at an increased computational cost and a larger set of free parameters ($2 d + 3$ for a $d$-dimensional problem). 

\subsection{The symmetric Hyperbolic likelihood function}
\label{seq:likelihood}

As shown in the earlier work of~\cite{Sasli:2023mxr}, a flexible likelihood formulation is the one of the symmetric Hyperbolic likelihood, where $\lambda = (d + 1)/2$ and only $\alpha$ and $\delta$ are left to control its overall shape. Here, we will follow the same methodology as in~\cite{Sasli:2023mxr}, but adapt it to parameter estimation for stochastic signals. For data $\bm{x}_i \in \mathbb{R}^d, 1 \leq i \leq n$, we derive the corresponding symmetric Hyperbolic (\hyp{}) log-likelihood function ($\Lambda_{\cal H}$) as 
\begin{equation}
\begin{aligned} 
\Lambda_{\cal H}(\alpha, \delta) = & n\Bigg[ \frac{d+1}{2} \ln\left(\frac{\delta}{\alpha}\right) + \frac{1-d}{2} \ln(2\pi)  \\
    & - \ln(2\alpha)  - \ln \left[K_{(d+1)/2} (\delta\alpha)\right]\Bigg] - \alpha\sum^n_{i=1}\sqrt{\delta^2 + r_i},
\label{eq:hyp}
\end{aligned}
\end{equation} 
where we have defined the quantity $r_i\equiv\bm{x}_i^{\rm T} \Delta^{-1} \bm{x}_i$. 

 
The likelihood of Eq.~(\ref{eq:hyp}) can also be rewritten for frequency domain data. To do that, we first define the noise-weighted inner product between two real time series $a$ and $b$ as~\cite{PhysRevD.69.082005, PhysRevD.49.2658}
\begin{equation}
	\langle a | b \rangle = 4 \, \mathrm{Re}\int\limits_0^\infty \mrm{d}f \left[ \tilde{a}^\ast(f) C_n^{-1}(f) \tilde{b}(f) \right],
	 \label{eq:inprod}
\end{equation}
where $C_n(f)$ is the 1-sided noise \gls{psd}, while the tilde $(\,\tilde{ }\,)$ denotes the Fourier transform, and the asterisk $(\,^\ast\,)$ the conjugate operation. For the case of multiple time series, or in other words multiple data channels, $C_n$ takes the form of a matrix, with the off-diagonal terms being the cross-spectral densities of the noise between the different data channels~\cite{Sasli:2023mxr}. Then, the residuals $r_i$ in frequency domain that enter eq.~(\ref{eq:hyp}) are given by   
\begin{equation}
r_i = {\rm Re}\, \left\{ \tilde{x}_i^\ast C_n^{-1} \tilde{x}_i \right\}. 
\label{eq:residualsf}
\end{equation}
Now, in frequency domain we need to carefully set the dimensionality $d$ of the problem (see eq.~(\ref{eq:hyp})), since the data for each channel are now assumed to be the sum of two distributions, one corresponding to the imaginary and one to the real part of the $\tilde{x}$ frequency series. Thus, we need to set $d=2n_\mathrm{c}$, where $n_\mathrm{c}$ the total number of data channels. In the time domain $d\equiv n_\mathrm{c}$, as already defined in eq.~(\ref{eq:hyp}).

The above eqs.~(\ref{eq:residualsf}) and~(\ref{eq:inprod}) are common in the literature, as they are used in Gaussian and Whittle likelihoods usually adopted in \gls{gw} data analysis, or in other applications in signal processing. In particular, the Whittle approximation of the likelihood is preferred when aiming to analyze stochastic signals, because the overall noise term $C$ depends on the models' parameters through $C (\param) \equiv C_n (\param_n) + C_i (\param_i)$, where $C_n (\param_n) [ C_i (\param_i)]$ denotes the noise (signal) model, and $\param=\param_n\cup\param_i$. This parameterization has been used in a plethora of studies concerning searches of stochastic signals~(see for example \cite{Cornish:2020dwh, LISACosmologyWorkingGroup:2022jok,Flauger:2020qyi,Contaldi:2020rht,Armano:2018ucz}). In that case, the logarithm of the likelihood reads
\begin{equation}
    \Lambda_{\cal W}(\param) \propto - \sum_f  \left( \ln C (\param)  + \langle x | x \rangle \right), 
\label{eq:whittle}
\end{equation}
As shown in the literature~\cite{whittle, choudhuri, roever2011B, roever2011A, vitale14}, the Whittle likelihood is approximate for Gaussian and stationary time series. This means that for a single channel and for the simplified scenario of residuals with the same statistics across all frequencies, the \hyp{} distribution will converge to the distribution of eq.~(\ref{eq:whittle}) if $\delta,\,\alpha \rightarrow \infty$, in such a way that $\delta / \alpha \rightarrow C (f)$, for the \gls{psd} $C(f)$ in eq.~(\ref{eq:residualsf}).

In~\cite{tang2021, 10.1214}, it was shown that the Whittle likelihood is still valid for the analyses of non-Gaussian time series, under some assumptions about the measured time series. One of these assumptions is that a relatively large sample size must be available, a requirement that cannot always be satisfied in realistic applications. That, in addition to high auto-correlation of the time series, might reduce the efficiency of this particular likelihood model. For a complete discussion on the validity of the Whittle approximation, we refer the reader to~\cite{Sasli:2023mxr, choudhuri, roever2011B, contreras}. Finally, we should also note here that the Whittle model needs to be transformed accordingly, depending on the windowing and averaging of the data, and in order to account for correlations between frequencies~\cite{Hindmarsh:2024ttn, Baghi:2023qnq, Pieroni:2020rob}.

In order to offset some of the potential disadvantages of any model that makes assumptions about the Gaussianity of the data, one might choose to adopt a likelihood model with tails heavier than those of the Gaussian distribution. This solution was proposed in~\cite{roever2011A, roever2011B}, where a likelihood was introduced based on the \stt{} distribution. In~\cite{roever2011A, roever2011B}, the authors adopted a prior for the variance of the noise that follows the $\mrm{Inv}-\chi^2$ distribution for each data point, which is a conjugate prior to the Gaussian likelihood. Then, marginalizing the noise variance yields the Student's t-distribution, which has the desired heavier tails compared to the Gaussian one. Other works explored similar ideas by combining multiple distributions to form the desired heavy-tailed likelihood~\cite{PhysRevD.60.021101, PhysRevD.65.122002}.

We follow a similar strategy in this paper, by using the hyperbolic likelihood model, with which, as shown in~\cite{Sasli:2023mxr}, we can approximate a large selection of exponential distributions. We will use the \hyp{} likelihood in order to extract stochastic signals and simultaneously estimate any potential departures from Gaussianities. The latter is done in practice by computing the $\xi$ parameter as~\cite{Prause1999TheGH,KUCHLER19991}
\begin{equation}
\xi = (1 + \alpha\delta)^{-1/2}, 
\label{eq:xi}
\end{equation}
which is scale and location-invariant, while it always holds that $0 \leq \xi < 1 $. In the previous relation we have assumed symmetrical distributions ($\beta\equiv0$). A value of $\xi$ close to zero indicates that our likelihood of \hyp{} has shaped to a Gaussian likelihood, which implies the Gaussianity of the underlying residual data. On the other hand, for $\xi\rightarrow1$ we get heavier-tailed distributions, indicating that the residual data depart from Gaussianity. Therefore, by estimating the parameters of the \hyp{} likelihood, together with the model parameters, we get additional information about the statistical properties of the residual data (see eq.~(\ref{eq:residualsf})). In the following sections, we will demonstrate this possibility by building a framework to estimate $\xi(f)$, which is frequency dependent, allowing us to probe the Gaussianity at different parts of the frequency spectrum of the data. 

We can visualize the distribution yielded by $\xi$, by plotting it on the so-called {\it shape triangle}. The shape triangle is a simple visualization tool, that allows us to easily categorize a given distribution given it's symmetry, skewness, and tailedness. An example is shown in the Appendix~\ref{sec:appendix}, see figure~\ref{fig:toys}. More details about the shape triangle can be found in~\cite{Sasli:2023mxr, Prause1999TheGH,KUCHLER19991,sed}

Finally, in a Bayesian analysis scheme, the posterior of the parameters \param{} is written as 
\begin{equation}
    p(\param | x) \propto \pi(x|\param)p(\param),
    \label{eq:bayes}
\end{equation}
where $ \pi(x|\param)$ is our chosen model for the likelihood, and $p(\param)$  the prior of the parameters. The marginal likelihood $\pi(x)$ usually acts as a normalization constant and for that reason is omitted from Eq.~(\ref{eq:bayes}), but it is useful for model selection purposes~\cite{gelmanbda04}.   

For the analyses in this work, we used \gls{mcmc} methods in order to sample all likelihoods (or posteriors), and estimate the properties of the given noise or stochastic signal. In particular, depending on the complexity of the problem at hand, we either used simple Metropolis-Hastings algorithms~\cite{mhsampler}, or the affine invariant sampler implementation \textsc{Eryn}~\cite{Karnesis:2023ras}. The former was used in Section~\ref{sec:toys}, where we draw $2\times10^5$ of samples, after a burn in period of $10^5$ samples. The Affine Invariant sampler, which was used mainly for the more demanding runs of Section~\ref{sec:lisa}, was configured with $140$ walkers and $5$ temperatures for the parallel tempering part of the algorithm~\cite{Karnesis:2023ras}. 

\section{Demonstration with simple examples
\label{sec:toys}}

In this section, we adopt a Bayesian framework to demonstrate the usage and performance of the \hyp{} likelihood in various simplified data analysis problems with synthetic data. First, we show a very simple application in Section~\ref{sec:toy_gaussian_white}, which focuses on estimating the variance of white Gaussian noise. Then, we extend the problem to non-Gaussian data in Section~\ref{sec:toy_nongaussian_white}. Finally, in Section~\ref{sec:lisa} we extend the analysis framework based on \hyp{}, defining a methodology to perform tests for departures for Gaussianity across the frequency spectrum.

\subsection{Gaussian and White noise
\label{sec:toy_gaussian_white}}

\begin{figure}
     \centering
     \begin{subfigure}[b]{0.52\textwidth}
         \centering
         \includegraphics[width=\textwidth]{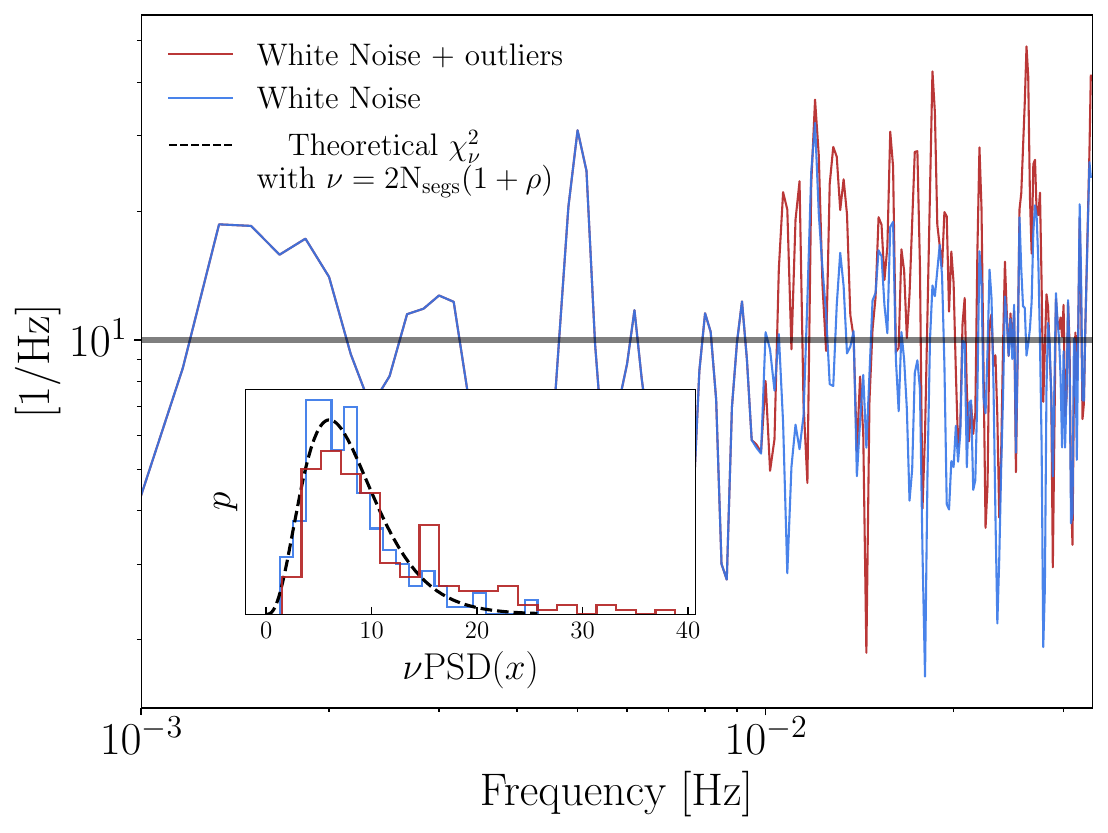}
         \caption{}
         \label{fig:toy_data}
     \end{subfigure}
     \quad\quad
     \begin{subfigure}[b]{0.4\textwidth}
         \centering
         \includegraphics[width=\textwidth]{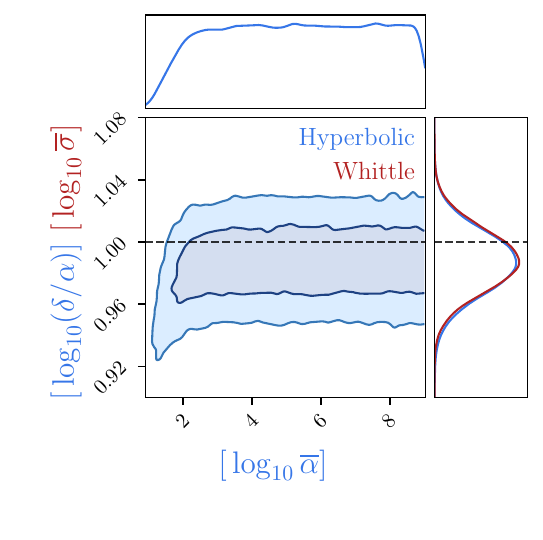}
         \caption{}
         \label{fig:toy_posteriors_gaussian_white}
     \end{subfigure}
        \caption{\textit{(a)} The synthetic data set generated for the application of Sections~\ref{sec:toy_gaussian_white} and ~\ref{sec:toy_nongaussian_white}. In Section~\ref{sec:toy_gaussian_white} we analyze the Gaussian data (blue), while in Section~\ref{sec:toy_nongaussian_white} the non-Gaussian data (red) which contain outliers. The averaged \gls{psd}s shown here, are computed with $N_\mathrm{segs}=4$ non-overlapping segments ($\rho=0$), and a rectangular window. The distribution of the \gls{psd} of the Gaussian data shown in blue, should follow a $\chi^2_\nu$ distribution with $\nu=8$ degrees of freedom (shown in the embedded panel with the dashed black line). \textit{(b)} Posteriors of the $\param=\{\log_{10}\delta/\alpha,\, \log_{10}\baralpha\}$ parameters of the likelihood \hyp{}, overlayed on the $\overline{\sigma}$ parameter recovered with $\Lambda_\whtl$. The black dashed line represents the true value of the level of the \gls{psd} of the Gaussianly distributed noisy data (see the blue curve in the left panel). The two distributions appear to be in excellent agreement. This, together with a very low figure for the Gaussianity index parameter $\xi$, shows that the \hyp{} likelihood has converged to the underlying Gaussian distribution of the data (see the text for details).}
        \label{fig:toy}
\end{figure}

We begin with a simple application to characterize the standard deviation of the Gaussian white noise. We simulate 1000 data points from ${\mathcal N}(0,\sigma)$ with $\sigma\simeq0.58$, generating time series with cadence $\Delta t = 15~\mrm{sec}$ (see figure~\ref{fig:toy_data}). We Fourier transform the data, so that the corresponding real and imaginary parts remain normally distributed. Given the above, we can adopt the Whittle likelihood of eq.~(\ref{eq:whittle}), and directly fit the overall level of the \gls{psd} of the data, a parameter that has a true value of $\overline{\sigma} = 2 \Delta t \sigma^2 = 10$. An alternative strategy would be to adopt the hyperbolic likelihood of eq.~(\ref{eq:hyp}) and fit the parameters $\alpha$ and $\delta$, as proxy for the noise level and the overall shape of the associated distribution. In fact, one can choose to fit a different but convenient combination of the $(\alpha,\delta)$ parameters for the \hyp{} likelihood. For example, since we know that for Gaussian data, the ratio between them $\delta/\alpha$ should converge to the unknown variance of the noise, we could choose it directly as one of the free parameters to be estimated (together with $\alpha$ or $\delta$, or any other combination of them). For this example, we fit the $\param=\{\log_{10}\delta/\alpha,\, \log_{10}\baralpha\}$, where $\baralpha = \alpha\delta$. We use broad uniform priors for both parameters $\log_{10}\baralpha\sim\mathcal{U}[0,10]$ and $\log_{10}\delta/\alpha\sim\mathcal{U}[-2,\,2]$. As already mentioned, we use \gls{mcmc} methods in order to obtain samples from the posterior and get estimates of $\param$.

\begin{figure}[!htb]
\floatbox[{\capbeside\thisfloatsetup{capbesideposition={left,center},capbesidewidth=.6\textwidth}}]{figure}[\FBwidth]
{\caption{Histograms of posterior medians obtained from $1000$ realizations of the toy-model Gaussian data as described in Section~\ref{sec:toy_gaussian_white}, for both the Whittle and the Hyperbolic likelihoods. In the limit  $\alpha,\,\delta\rightarrow\infty$ such as that $\delta/\alpha\rightarrow\sigma^2$, the two likelihoods yield a similar distribution, as expected.}\label{fig:toy-mc}}
{\includegraphics[width=.3\textwidth]{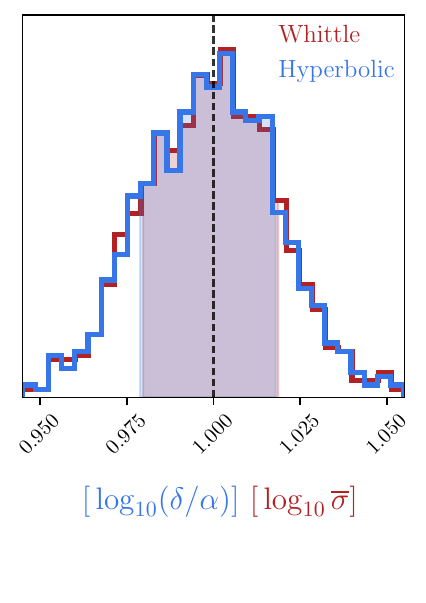}}
\end{figure}

The results are displayed in figure~\ref{fig:toy_posteriors_gaussian_white}, where the posteriors for the logarithm of the $\delta/\alpha$ parameter for $\Lambda_\hyp$, and $\overline{\sigma}$ for the $\Lambda_\whtl$ likelihoods are shown. As expected, since the data follow a Gaussian distribution, the recovered posteriors of the two parameters show excellent agreement (see Figure~\ref{fig:toy_posteriors_gaussian_white}). With the parameter estimates of $(\alpha,\delta)$ we can also get a measure for the Gaussianity of the given data, by directly computing the $\xi$ parameter from eq.~(\ref{eq:xi}). We get a value very close to zero ($\xi=0.002$), which verifies that the distribution of the residual data follows the normal distribution. In order to validate this result, we also generated $1000$ realizations of the noise and repeated the analysis for both likelihoods. The histogram of the median of the posterior of each realization is shown in Figure~\ref{fig:toy-mc}. The two distributions appear to be almost identical and centered around true value, as expected.

\subsection{Non-Gaussian noise
\label{sec:toy_nongaussian_white}}

In this example, we extend the application of the \hyp{} model to non-Gaussian data. We use the previously generated data set (white Gaussian noise), and add a set of sinusoidal signals, with random frequencies and amplitudes. In particular, we inject twenty signals with their amplitude drawn from a uniform in $\log_{10}$ ($\mathcal{U}[-2, -1]$), and their frequency between $0.01$ and $0.05~\mathrm{Hz}$. That allows us to simulate a certain segment in frequency, which does not have Gaussian properties any more. In particular, the data below $10~\mathrm{mHz}$ follow the Gaussian distribution as defined in the previous subsection~\ref{sec:toy_gaussian_white} (see figure~\ref{fig:toy_data}, the blue and red curves overlap for frequencies below $10~\mathrm{mHz}$), while above that limit the statistics of the data have changed to an unknown distribution shape. We thus split the frequency domain into two segments, above and below $10~\mathrm{mHz}$, and extend eq.~(\ref{eq:hyp}) to fit independent parameters in each segment. Following the strategy of the previous example, we now have two pairs of $\param_{\hyp, j}=\{\log_{10}\delta/\alpha,\, \log_{10}\baralpha\}_j$, where $j=\{ \mathrm{low},\, \mathrm{high}\}$ identifies the two segments in frequency. We do the same with the Whittle likelihood of eq.~(\ref{eq:whittle}), fitting for the parameters $\param_\whtl=\{\log_{10}\overline{\sigma}_\mathrm{low}, \log_{10}\overline{\sigma}_\mathrm{high} \}$.

\begin{figure}[!htb]
     \centering
     \begin{subfigure}[b]{0.45\textwidth}
         \centering
         \includegraphics[width=\textwidth]{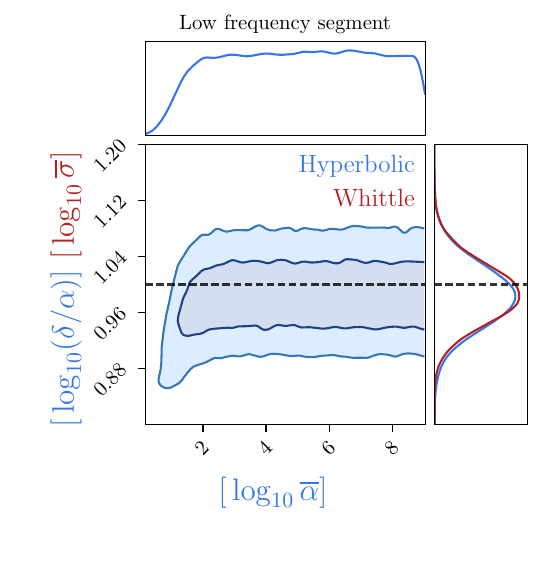}
         \caption{}
         \label{fig:toy_posteriors_nongaussian_seg1}
     \end{subfigure}
     \quad\quad
     \begin{subfigure}[b]{0.45\textwidth}
         \centering
         \includegraphics[width=\textwidth]{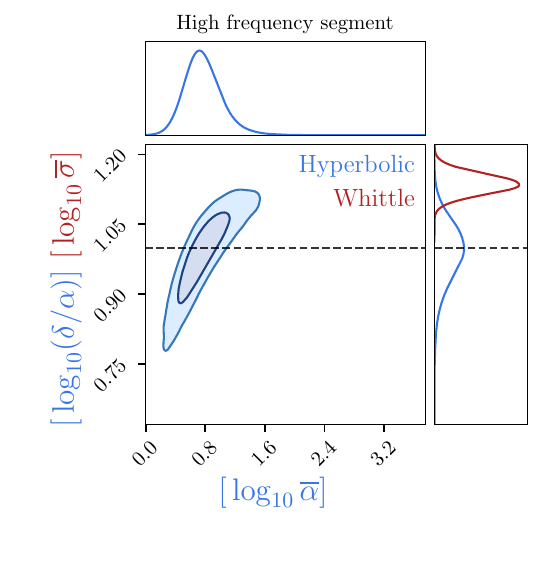}
         \caption{}
         \label{fig:toy_posteriors_nongaussian_seg2}
     \end{subfigure}
        \caption{\textit{(a)}  Posteriors of the parameters of the \hyp{} likelihood for the low frequency segment of the data, which follow the normal distribution. As expected, the $\delta/\alpha$ parameter converges to the actual injection parameter for the variance of the unknown noise. The posterior for the variance of the data as sampled with the Whittle likelihood is also shown in dark red colour, in order to highlight the agreement between the two methodologies for Gaussian data. The true value is indicated with the black dashed line. \textit{(b)} The same for the high frequency segment considered in the analysis, where the data do not follow Gaussian statistics. The Whittle likelihood yields biased results for the variance of the data, whereas the \hyp{} likelihood allows us to better probe the statistical properties of the synthetic data set by being more flexible than the whittle approximation of the likelihood.}
        \label{fig:toy_posteriors_nongaussian}
\end{figure}

We again sample the parameters using \gls{mcmc} methods and get the posteriors, which are presented in figure~\ref{fig:toy_posteriors_nongaussian}. For the low-frequency segment (see left panel \ref{fig:toy_posteriors_nongaussian_seg1}), it is clear that $\Lambda_\whtl$ and $\Lambda_\hyp$ recover almost identical posteriors for the noise level. This is expected, given the theoretical predictions and the demonstration from the previous example. The situation changes for the high-frequency segment (figure~\ref{fig:toy_posteriors_nongaussian_seg2}, right panel), where the data do not follow the normal distribution. In fact, we notice that the Whittle likelihood appears to yield biased results for the level of the noise \gls{psd}, which can be attributed to the fact that it adopts erroneous assumptions about the statistical properties of the noise. On the other hand, the model based on the hyperbolic distribution yields a result indicating deviation from the normal distribution. This can be verified as in the previous example, by computing the parameter $\xi$ for the two frequency segments, which returns $\xi_\text{low} = 0.0037$ and $\xi_\text{high} = 0.80$ respectively. Thus, we obtain quantitative confirmation of the statistical properties of the noise for the two segments considered in the analysis. This analysis strategy shows great potential for the analysis of data from \gls{gw} observatories, where the instrumental noise is not well known a priori for the complete frequency range (see Section~\ref{sec:lisa}). 
 
\section{Applications to Gravitational Wave data
\label{sec:lisa}}

Having introduced a basic analysis framework based on the $\Lambda_\hyp$ likelihood, we can go one step further and apply it in to more demanding scenarios, directly drawn from the future \gls{lisa} mission. As discussed above, the \gls{lisa} data set is expected to be populated with different kinds of \gls{gw} signatures, both transient and monochromatic, as well as stochastic signals~\cite{Colpi:2024xhw}. We will focus on two example applications. For the first case, we will examine the case of a colored Gaussian cosmological signal in colored Gaussian noise, while for the second case, we will investigate possible departures from Gaussianity of the ``confusion'' foreground signal generated by the unresolved \gls{ucbs}. In both cases, we will need to employ a methodology to probe the properties of the residual data across the spectrum.

\subsection{Detecting stochastic signal in colored noise: First-order phase transitions
\label{sec:lisa_fopt}}

\begin{figure}
    \centering
    \includegraphics[width=.9\textwidth]{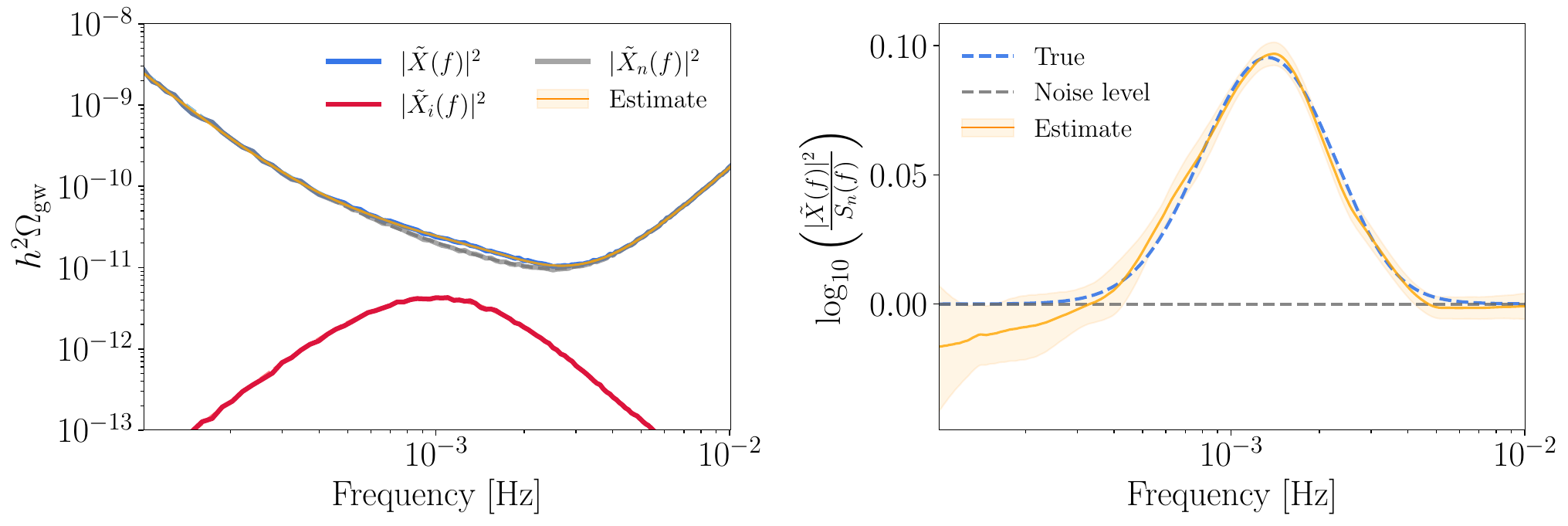}
    \caption{\textit{Left}: Breakdown of the synthetic data set generated for the application of Section~\ref{sec:lisa_fopt}. The blue curve represents the measured data, while the gray and red the noise and signal respectively. The dashed lines represent the ground truth. \textit{Right}: Representation of the result for the excess of power due to the presence of a stochastic signal in the data. Here we plot the logarithm of the deviations from the assumed spectral model of the noise (gray dashed line), while the true total excess of power is represented with the blue dashed line. The shaded yellow area shows the 2$\sigma$ confidence interval for the recovered spectral shape of the data, using the \hyp{} likelihood and the methodology explained in the text.}
    \label{fig:phase-trans_search_data}
\end{figure}

We begin with the first application, where our aim is to search for a stochastic signal in coloured noise. We choose to simplify the procedure, and thus use simple analytical models for the power spectra, which we use to simulate the data directly in the frequency domain (see left panel of figure~\ref{fig:phase-trans_search_data}). The noise \gls{psd} has a shape following the LISA sensitivity in $\Omega_\mathrm{gw}$ units~\cite{Babak:2021mhe}. The signal $X_i(t)$, which in our analysis we consider unknown, is generated from a First-order Phase-transitions cosmological model~\cite{LISACosmologyWorkingGroup:2022jok}. 

This type of signal, as the name suggests, originates from bubbles of First-Order \gls{pt} colliding during the early epochs of the Universe, prior to or during the Big Bang nucleosynthesis. For some physical models, these collisions produce a potentially detectable stochastic \gls{gw} signal in the $~\mathrm{mHz}$ range~\cite{Caprini:2019egz, weir_2022_6949107}. Following the formalism of~\cite{weir_2022_6949107}, the spectral shape of this signal can be expressed as
\begin{equation}
    h^2\Omega(f) = h^2\Omega_\mathrm{sw}\left(\frac{f}{f_{\mathrm{p},0}}\right)^3 \left[\frac{7}{4+3\left(\frac{f}{f_{\mathrm{p},0}}\right)^2}\right]^{\frac{7}{2}} 
    + h^2\Omega_\mathrm{turb} \frac{\left(\frac{f}{f_\mathrm{turb}}\right)^3}{\left[1+\left(\frac{f}{f_\mathrm{turb}}\right)\right]^{\frac{11}{3}} \left(1+\frac{8\pi f}{h_*}\right)},
    \label{eq:fopt}
\end{equation}
where $f$ the given frequency vector, and $h=0.67$. The stochastic signal can be divided into two additive terms, one corresponding to the sound wave component~\cite{Caprini:2019egz}, and one to the turbulence part of the process~\cite{Weir:2017wfa}. During the first phases of the \gls{pt} process, the different states of vacuum are separated by a barrier that will eventually cross the barrier through
bubble nucleation~\cite{Caprini:2024hue}. As the shells of kinetic energy from different bubbles collide and overlap, \gls{gw}s are produced due to the propagation of the kinetic energy of the fluid in the primordial plasma, which travel at the speed of sound~\cite{Caprini:2024hue}. This process corresponds to the sound wave part of the expression~(\ref{eq:fopt}). After this first acoustic phase, a turbulent regime might rise, which could potentially produce detectable components of the stochastic signal~\cite{Weir:2017wfa,Caprini:2024hue}. This process is described by the second term of eq.~(\ref{eq:fopt}). The $\Omega_\mathrm{sw}$ and $\Omega_\mathrm{turb}$, are amplitude parameters for those processes, respectively. These amplitudes and the $h_*$, $f_{\mathrm{p},0}$, and $f_\mathrm{turb}$, are constants that depend on the physical assumptions of the process. The resulting spectrum takes the form of a bump with a tail towards the higher frequencies, as shown in the left panel of figure~\ref{fig:phase-trans_search_data}. For more details on the description of the physical process of the First-Order \gls{pt} and the generated \gls{gw} signal, we refer the reader to~\cite{LISACosmologyWorkingGroup:2022jok,Caprini_2018,Caprini:2019egz,Caprini:2024hue,weir_2022_6949107,Weir:2017wfa}. Finally, we generate data for three years, with a cadence of 15~$\mathrm{sec}$. We also assume that this is the only signal that the observatory measures, which is an unrealistic assumption~\cite{LISACosmologyWorkingGroup:2022jok}, but it is acceptable for the purposes of this exercise.

This problem, due to the colored noise and the signal, requires more careful modeling than the first two toy example applications, because the dependence on the frequency $f$ should be taken into account. However, fitting for the \gls{psd} or the ratio $\delta/\alpha$ for each frequency point is unrealistic, especially for long measurements. Thus, it is necessary to adopt a proxy function instead, being both flexible across the frequency range, and  minimizing the number of free parameters for inference. Given the above, a shape-agnostic spectral model based on B-spline interpolation is ideal for this job (more specifically, we use the Akima interpolation method~\cite{Akima1970ANM}). We choose to follow a similar procedure as in~\cite{Baghi:2023qnq}, where we first define a set of spline knots equally spaced in log-frequency, and consider the spline knots amplitudes as free parameters to be estimated. The quantity to be interpolated is the deviations from the the \gls{psd} level of the data for the case of the Whittle likelihood, or the $\delta/\alpha$ ratio for the hyperbolic one (which again, for Gaussian data is expected to show deviations from our assumed \gls{psd} model). In the end, with the above methodology, we have moved a step further from splitting the data into segments and estimating the individual parameters at each segment as done in the previous application. 

In order to allow for a more dynamical range, we do not infer on $(\delta/\alpha)$ directly, but instead interpolate on its logarithm as $\log_{10}(\delta/\alpha)(f)$, by fitting the spline knots amplitudes parameters $\kappa_{\delta/\alpha,i}$, where $i$ the given knot number for $f_i$. For the sake of simplicity, we consider that $\baralpha(f)\equiv\baralpha$ is the same for all frequencies, which means that the non-Gaussianity indicator applies to the complete spectrum. In our next application this choice will be relaxed as well. 

We choose to fit for the amplitudes of five equally-spaced in log spline knots to detect any deviations from the assumed model of the noise $S_n$. As in previous sections, we use \gls{mcmc} methods to sample the posteriors of the knots amplitudes. We choose as priors  for the knot amplitudes $\kappa_{\delta/\alpha,i}\sim\mathcal{U}[-2, 4]$, and for the \baralpha{} parameter $\log_{10}\baralpha\sim\mathcal{U}[-1, 20]$. It is worth to mention here, that one could allow for more flexibility, by using dynamical samplers and estimate both the parameters of the spline knots and their optimal number simultaneously~\cite{Baghi:2023qnq,Karnesis:2023ras}.

The result is shown in figure~\ref{fig:phase-trans_search_data}. The logarithm of the ratio of the deviations from the assumed model to the true data is shown together with the estimate of the fit.  The baseline, to which we estimate deviations from, is shown with the dashed gray line, while the full measurement data (dashed, blue), is deviating from the baseline due to the presence of an unknown stochastic signal. The flexible spline model is capable of characterizing this deviation, by yielding a $\delta/\alpha$ depending on the frequencies, following this excess of power due to the stochastic signal component $X_i(t)$. The combination of the $(\delta, \alpha)$ parameters shows that the data are Gaussianly distributed, as expected. This can be verified by computing the value of $\xi$ in the frequency range of the analysis. We compute $\xi(f)=0.002\pm0.003$ compatible with zero for each spline knot parameter, which is in agreement with the setup of this simulation. 

\subsection{Characterizing the astrophysical foreground signal of Ultra Compact Binaries 
\label{sec:lisa_ucbs}}

\begin{figure}
    \centering
    \includegraphics[width=.65\textwidth]{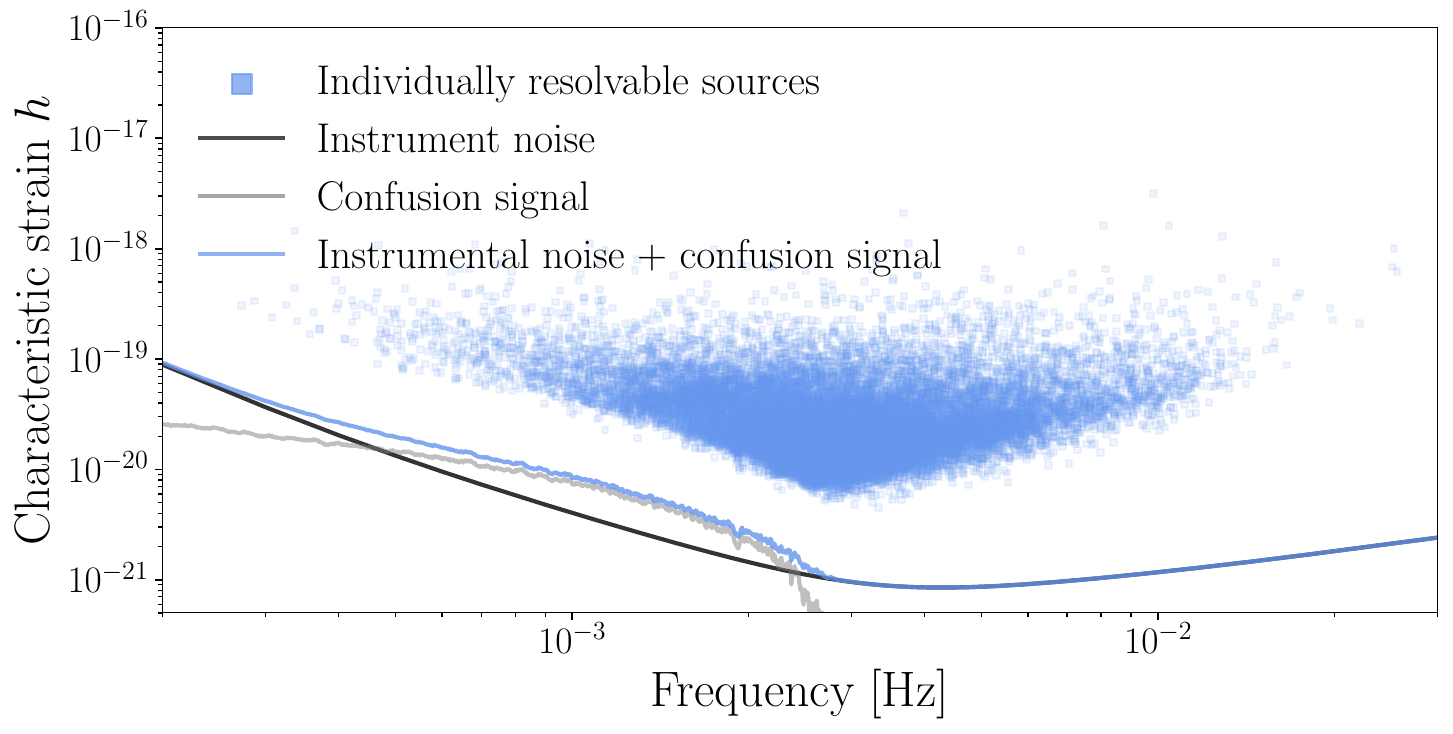}
    \caption{Characteristic strain of the signal induced by the \gls{ucbs}. The square points represent the sources that are individually resolvable, while the gray line the confusion signal generated by the unresolved binaries. The blue line represents the overall measured data, as the sum of instrumental noise (black) plus the aforementioned confusion signal. The resolved binaries and confusion signal were estimated with the methodology resented in~\cite{PhysRevD.104.043019}, for a mission duration of $T_\mathrm{obs}=4~\mathrm{years}$. Here, the non-stationarity of the confusion signal is not represented, because all the data in this figure are derived from the \gls{psd} of the data of the complete duration of the measurement.}
    \label{fig:galaxy}
\end{figure}

We now apply this methodology to a more elaborate example application, which focuses on the ensemble signal of the \gls{ucbs} in the vicinity of our Milky Way. The orbital evolution of this type of binaries, which is driven by \gls{gw} emission, is really slow compared to the overall observation time of the future \gls{lisa} mission. For this reason, they emit an almost monochromatic \gls{gw} signal, which appears as spikes in the frequency data series~\cite{Strokov:2023ypy}. \gls{lisa} will measure the majority of the \gls{gw} signals from the \gls{ucbs} that are distributed in the Galaxy ($\mathcal{O}(10^7)$), while only a small percentage of those will be resolvable. Depending on the given population model, around $\mathcal{O}(10^4)$ objects~\cite{LISACosmologyWorkingGroup:2022jok, Colpi:2024xhw} will be well measured from the data. The rest will generate a non-stationary confusion stochastic signal, which might also have non-Gaussian properties (see figure~\ref{fig:galaxy}). The non-stationarity is caused by the orbital motion of the \gls{lisa} constellation, pointing in and out the galactic center, where the density of those objects rapidly increases~\cite{Colpi:2024xhw, Korol:2020lpq, Georgousi:2022uyt}. 

Understanding the statistical properties of this contribution is crucial for extracting the full scientific potential of the future \gls{lisa} mission, simply because the recovery of any \gls{gw} signature will depend on our level of understanding of the underlying total noise. This means that testing for the statistical properties of the spectrum at all frequencies is necessary. Here, we will do that by separating the data in small frequency bins and fitting for the $\alpha$ and $\delta$ (or a suitable reparametrization) for each of them. The procedure is described in the following.

\subsubsection{Using the measurement as a single segment \label{sec:galaxy_signle_segment}}

For our purposes, we simulate the \gls{gw} signals of all binaries for $T_\mathrm{obs} = 4~\mathrm{years}$, and instrumental noise directly from the theoretical \gls{psd}~\cite{Colpi:2024xhw}. The Galaxy population model is the same used in previous works, mostly focusing on \gls{lisa} data analysis challenges~\cite{Colpi:2024xhw,Karnesis:2023ras,Korol:2020lpq,Nelemans:2013yg,PhysRevD.104.043019,Katz:2024oqg,Littenberg:2023xpl,Strub:2024kbe}. The first step is to generate the \gls{ucbs} signals directly in the frequency domain using the \textsc{FastGB} waveform software~\cite{Cornish:2007if}. We then apply the iterative procedure of~\cite{Karnesis:2023ras} (which was first introduced in~\cite{2006PhRvD..73l2001T}), in order to estimate both the number of resolvable sources, as well as the residual confusion signal generated by the remaining unresolved sources. The iterative procedure is based on a \gls{snr} criterion, where sources that produce an \gls{snr} above a chosen threshold are classified as resolvable and are subtracted exactly from the data. After subtraction at each iteration, the total noise, which is the sum of the instrumental noise and the confusion \gls{gw} signal, is re-estimated and the classification process repeats. When convergence is reached, i.e. no more sources are identified as resolvable, we get a final estimate of the catalog of the sources that will generate the  the confusion \gls{gw} signal. This algorithm was further improved in~\cite{Digman:2022jmp}, in order to account for the time variation of the signal. 

Then, we use this catalog of residual sources in order to generate the signal time series to be used in our analysis. For that purpose, we use the \textsc{Balrog} code, which is a suite of tools used to perform a variety of inferences on LISA sources and instrumental artifacts. This includes supermassive binary black hole mergers~\cite{2023PhRvD.108l4045P}, \gls{ucbs}~\cite{2019PhRvD.100h4041B,2020ApJ...894L..15R,2023MNRAS.522.5358F,2024MNRAS.531.2817M}, stellar-mass binary black holes~\cite{2021PhRvD.104d4065B,2022arXiv220403423K,2023PhRvD.108h4014B,2024arXiv240813170B}, glitches~\cite{2023PhRvD.108l3029S}, and GW stochastic backgrounds. The final \gls{psd} of the time series appears as a spectral ``bump'' between $0.3$ and $3~\mathrm{mHz}$, right at the most sensitive frequency region of the instrument~\cite{Colpi:2024xhw}, as expected (see Figure~\ref{fig:galaxy}). 

In order to demonstrate the methodology introduced here, we choose to work with the time series of a single \gls{tdi} channel~\cite{Tinto:2004wu}. In particular, we use the channel $\mathrm{A}$, one of the noise-orthogonal \gls{tdi} combinations~\cite{Tinto:2004wu,PhysRevD.66.122002}. In reality, it will be quite challenging to obtain true noise-orthogonal \gls{tdi} combinations, because this would require equal noise levels across the three satellites~\cite{Baghi:2023qnq,Muratore:2023gxh}, as well as minimal fluctuations of the shape of the triangular constellation in time~\cite{Hartwig:2023pft}. In a more realistic scenario, the full covariance matrix of all \gls{tdi} channels would be plugged into eq.~(\ref{eq:residualsf}). However, a single channel such as the \gls{tdi} $\mathrm{A}$ is suitable for the purposes of this exercise. 

To simplify the process even further, we whiten the data with a model that is computed numerically by applying a smoothing filter directly on the power spectrum of the data. In particular, we filter following a two-step process: we first apply a running median on the \gls{psd} of the data, then we further smooth it through nonparametric kernel regression~\cite{seabold2010statsmodels}. This practice offers certain advantages. Firstly, it allows us to fit for deviations from a given model instead of the actual spectral amplitude. Secondly, it allows us to perform a joint analysis between different data segments, while fitting for common parameters for all segments (this is done in the following Section~\ref{sec:cyclo_analysis}, where the cyclo-stationarity of the signal is accounted for). However, we should note here that it may introduce nuances in the interpretation of the results. Using the data to construct a whitening filter could suppress higher-order moments of the data, which are important for detecting deviations from Gaussianity. Nevertheless, in our case, we expect this effect to be minimal, as our whitening process is quite robust to outliers~\cite{PhysRevD.104.043019}. 

We will then use the hyperbolic likelihood of eq.~(\ref{eq:hyp}) in order to both fit the noise \gls{psd} levels, as well as any potential departures from Gaussianities. By assuming uncorrelated data-points in the frequency domain, we split the \gls{psd} data in bins that are evenly spaced in log-space. We use 20 bins in total, and we essentially apply the hyperbolic likelihood of eq.~(\ref{eq:hyp}) at each bin. Then, similarly to the first example of Section~\ref{sec:toy_nongaussian_white}, for each frequency segment we estimate the parameters $\param_{\hyp, j}=\{\log_{10}\delta/\alpha,\, \log_{10}\alpha\}_j$, where $j$ identifies a given frequency bin. As a reminder, we expect to get an estimate of the overall \gls{psd} level from the $\log_{10}\delta/\alpha$ parameters, as well as an estimate for the statistical distribution of the residuals by computing $\xi$ from eq.~(\ref{eq:xi}). As priors, we choose the uniform distribution with very broad bounds $\mathcal{U}[-10,\, 10]$, for all the parameters $\param_{\hyp, j}$. Then, the parameters for all segments are sampled in a joint likelihood scheme with \gls{mcmc} methods~\cite{Karnesis:2023ras}.

\begin{figure}
    \centering
    \includegraphics[width=\textwidth]{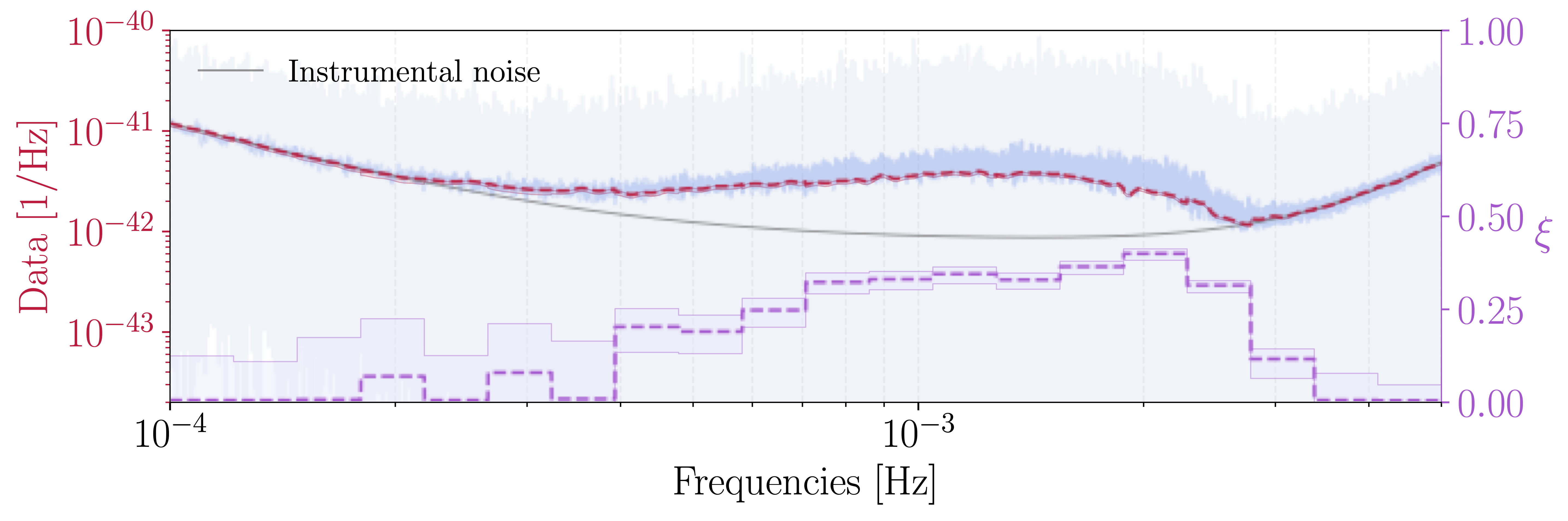}
    \caption{Result of the analysis presented in Section~\ref{sec:galaxy_signle_segment}. The left red $y$-axis corresponds to the spectral data in shades of blue and red. The light blue represents the \gls{psd} of the data, while the darker blue is a smoothed version plotted for reference. The dashed red line is the resulting \gls{psd} level from the MCMC analysis, plotted directly from the MCMC chains of the $\log_{10}\delta/\alpha$ parameter for each $j$ frequency bin. The 2$\sigma$ error bar is plotted as a shaded area as well, but it's quite small and not visible in this graph. The right $y$-axis corresponds to the $\xi$ parameter, which is represented by the dashed purple line. Values of $\xi>0$ indicate departures from Gaussianity. See text for more details.}
    \label{fig:result_galaxy}
\end{figure}

Results are summarized in figure~\ref{fig:result_galaxy}. The \gls{psd} of the data is shown in light blue, whereas a smoothed version of the same data is shown in darker color for reference. The theoretical instrumental noise is represented by the gray line. This deviation of the data from the gray line is all due to the confusion stochastic \gls{gw} signal by the unresolved \gls{ucbs}. The dashed red and purple lines represent the results of the fit, which are discussed in detail below.

\paragraph{Overall level of the noise} As already mentioned, an estimate for the overall \gls{psd} level of the data can be given by the $\log_{10}\delta/\alpha$ parameter for each of the segments in the analysis. This is exactly shown with the red dashed line in figure~\ref{fig:result_galaxy}. The $2\sigma$ error bars are also plotted as a transparent area, but are too small to be distinguishable in the plot. It is quite evident that the general shape of the total noise as the sum of the instrumental noise plus the stochastic signal has been recovered. On the other hand, it seems that the recovered \gls{psd} level is slightly lower than the \gls{psd} of the data (see darker blue line in figure~\ref{fig:result_galaxy}). This is true only in regions where the confusion stochastic \gls{gw} signal is dominant. At the edges of the spectrum the red dashed line matches exactly the instrumental noise. This divergence can be attributed to the cyclo-stationary nature of the confusion signal, which is mainly caused by the orbital position of the observatory~\cite{2024cycloPozzoli}. This means that with the proposed heavier-tailed likelihood framework we are able to compensate for this non-stationarity by adapting the overall shape of the likelihood. The final result converges to an overall \gls{psd} baseline which is lower than the one numerically computed with the Welch method at the frequency range where the stochastic signal is dominant. We should note here that one should be careful with the interpretation of the results for values of $\xi$ much larger than zero, because the residuals do not follow the normal distribution, and the $\delta/\alpha\rightarrow\sigma^2$ does not hold any more.

\paragraph{Statistical distribution of the residuals} The rightmost $y$-axis of figure~\ref{fig:result_galaxy} refers to the level of $\xi$, which is estimated as a function of frequency (purple dashed line). The $\xi$ parameter, which is given by eq.~(\ref{eq:xi}), is a measure of the taildness of the hyperbolic likelihood. Values close to $0$ indicate convergence to the Gaussian distribution, while values close to unity indicate convergence to the Laplace distribution~\cite{Sasli:2023mxr}. The $\xi(f)$ as shown in Figure~\ref{fig:result_galaxy}, clearly illustrates the parts of the spectrum where a heavier-tailed distribution is preferable to the Gaussian. As expected, this effect is dominant at frequencies where the stochastic \gls{gw} signal is prevalent. However, with this result alone, it is not possible to distinguish between non-stationarities or possible non-Gaussianities in the data. For example, we would expect to detect non-Gaussian regions in the spectrum, especially at frequencies where the stochasticity of the confusion \gls{gw} signal drops (between $2$ and $3~\mathrm{mHz}$ in the plot)~\cite{riccardo,robbieTysonSGWB_stub}. Any such non-Gaussian effect is probably subdominant to the effect of the cyclo-stationarity of the signal. For that reason, we need to proceed with an analysis that takes into account this bi-annual modulation.

\subsubsection{Accounting for the cyclo-stationarity of the galactic foreground}
\label{sec:cyclo_analysis}

\begin{figure}
     \centering
     \begin{subfigure}[b]{0.45\textwidth}
         \centering
         \includegraphics[width=\textwidth]{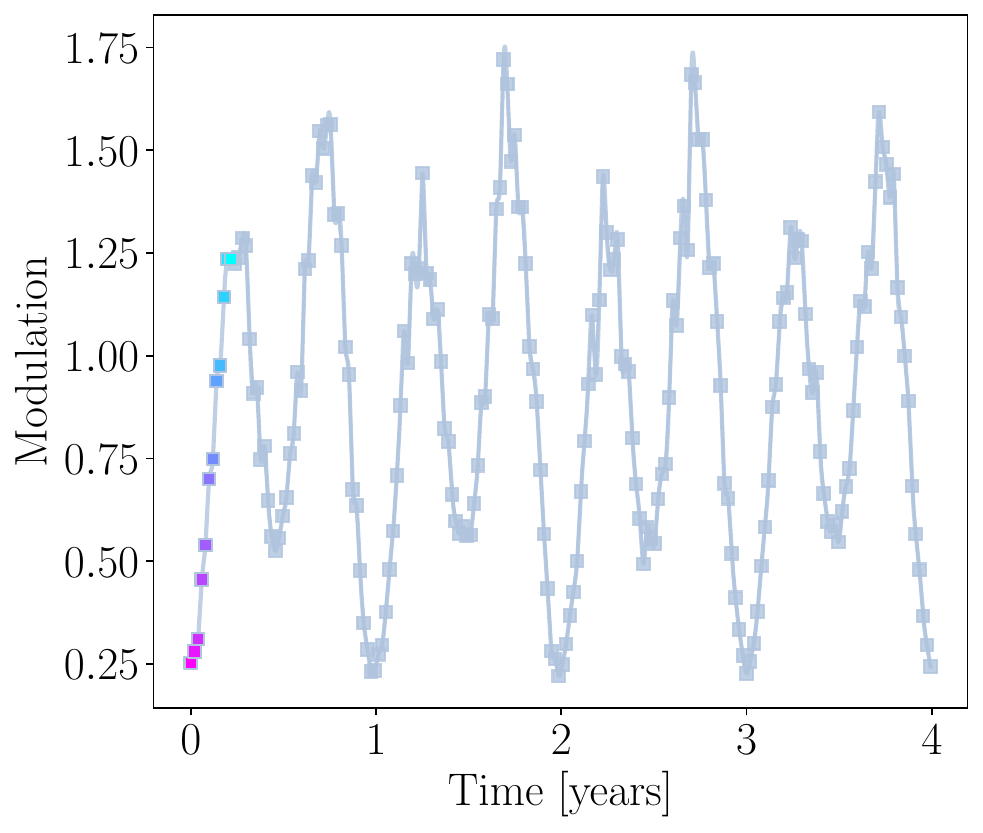}
         \caption{}
         \label{fig:modul_ratio}
     \end{subfigure}
     \quad\quad
     \begin{subfigure}[b]{0.45\textwidth}
         \centering
         \includegraphics[width=\textwidth]{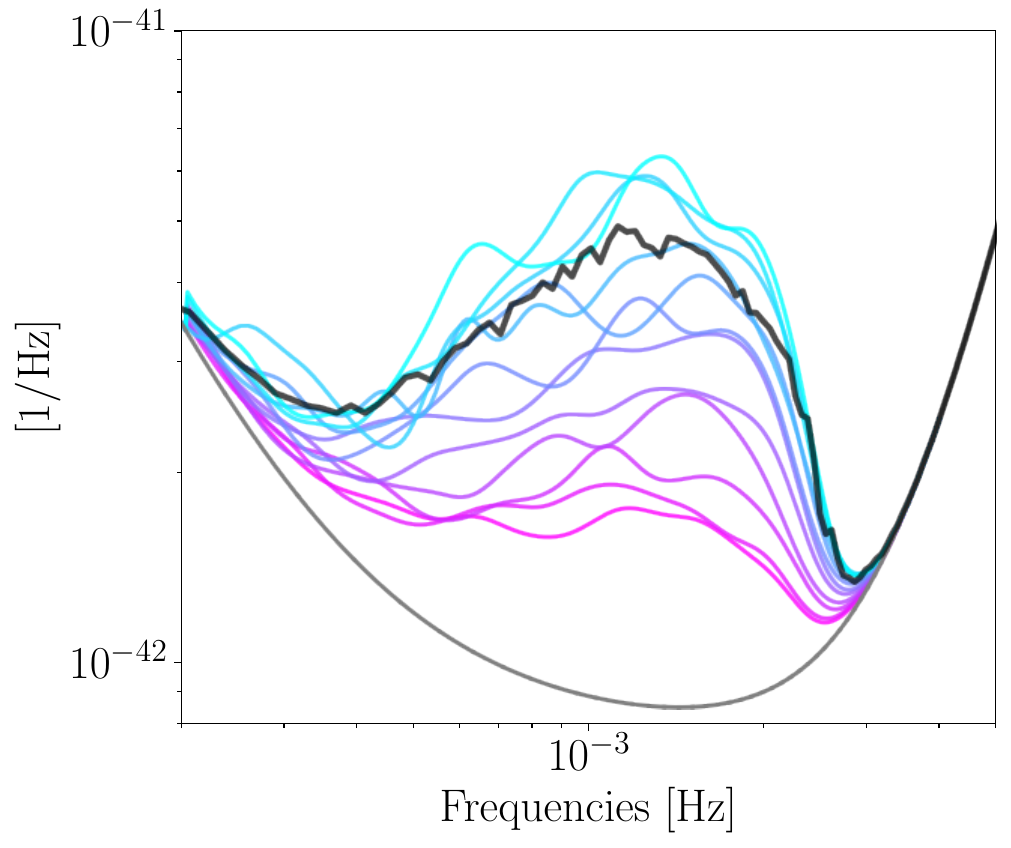}
         \caption{}
         \label{fig:modul_spec}
     \end{subfigure}
        \caption{\textit{(a)} Modulation of the stochastic \gls{gw} signal generated by the \gls{ucbs} as measured by \gls{lisa}. Each of the data points of the modulation is computed as the median (for all frequencies) of the ratio between the \gls{psd} of the confusion signal for the duration of the measurement, to the \gls{psd} of the confusion for a time segment of two weeks. The square points indicate the beginning of each of the $2$-week segments that were considered in the analysis. \textit{(b)} The (smoothed) \gls{psd} of the data for the first $12$ segments (the colour map denotes increasing segment number, see left panel \textit{(a)}), as a demonstration of the time modulation of the \gls{psd} of the confusion signal. The \gls{psd} or the complete duration is shown in black, while the instrumental noise model in gray color.}
        \label{fig:modulation}
\end{figure}

There have been several studies in literature focusing on the analysis of the \gls{lisa} data in the presence of a time-varying component of the noise (e.g. see~\cite{Digman:2022jmp, Boileau2021, Hindmarsh:2024ttn, Alvey:2024uoc,2024cycloPozzoli, Cornish:2020odn} and references therein). Here, we choose to follow the straightforward strategy of analyzing the data in short time segments and in a joint likelihood scheme. In particular, we split the time series data in $N_\mathrm{seg} = 202$ segments with an approximate duration of $1$~week each. The \gls{psd} of only the stochastic \gls{gw} confusion signal is then computed for each segment\footnote{As before, we have used the Welch method and the Blackman-Harris spectral window to compute the periodograms for each segment. The resulting \gls{psd} is then further smoothed by applying a running median of the \gls{psd} of the data, and Gaussian filters available with the {\tt scipy} library~\cite{2020SciPy-NMeth}.}, and compared to the \gls{psd} of the signal for the entire $4$-year duration of the measurement. Their ratio is presented in figure~\ref{fig:modulation}, which can be considered as a measure of the modulation and cyclo-stationarity of the overall stochastic signal.

Then, just as done for the previous Section~\ref{sec:galaxy_signle_segment}, the whitened power spectrum for each segment is computed, to be used as an input for $r_i$ in the likelihood of eq.~(\ref{eq:hyp}). In this simplified analysis scheme, the different segments are assumed to be independent, which essentially allows us to sum their corresponding likelihoods. The $\lambda$ parameter in eq.~(\ref{eq:hyp}) has to be adjusted as well, by injecting the appropriate $d=2N_\mathrm{seg}$ (see section~\ref{seq:likelihood}). By doing all the above, we expect to compensate for the modulation of the stochastic signal shown in figure~\ref{fig:modulation}, and thus investigate for potential non-Gaussian parts in the overall spectrum.   

\begin{figure}
    \centering
    \includegraphics[width=\textwidth]{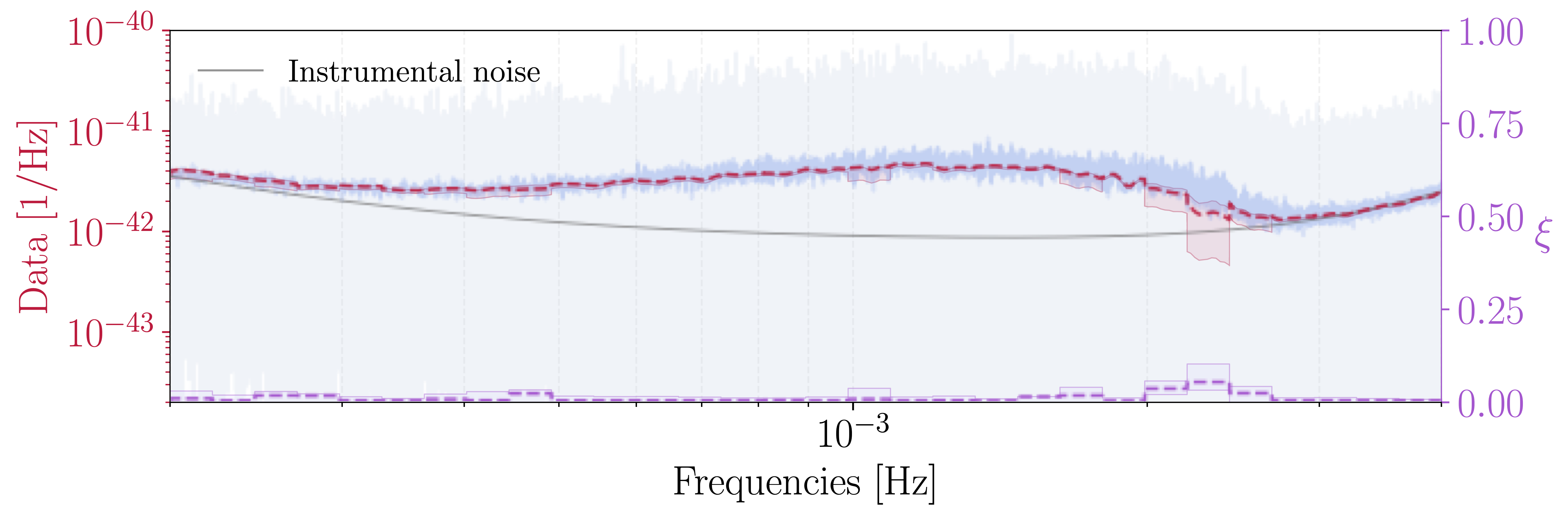}
    \caption{Same as in figure~\ref{fig:result_galaxy}, but for the demodulated scheme of fitting the data jointly, for a total of $N_\mathrm{segs}=202$ data segments, with an approximate duration of one week each. It is notable that most of the heavier tails due to the non-stationarity of the stochastic signal are now suppressed, but a detectable increase of $\xi$ between $2$ and $3~\mathrm{mHz}$ remains. This indicates a potential non-Gaussianity of the signal, which independently confirms the findings of~\cite{riccardo,robbieTysonSGWB_stub}. See the text for details. }
    \label{fig:result_galaxy_demod}
\end{figure}

We proceed by assuming common $\param_{\hyp, j}$ parameters for all data segments and their $j$ frequency bins. The rest of the parameter estimation procedure is identical to the previous section, assuming the same broad priors and sampling the $\param_{\hyp, j}$ with MCMC methods. The results are shown in figure~\ref{fig:result_galaxy_demod}. It is quite evident, that the $\xi(f)$, which measures any given departures from Gaussianity, now remains compatible with $0$ within the errors, for almost all frequencies. This means that the effect of the cyclo-stationarity of the stochastic signal is almost entirely suppressed by the analysis strategy that we followed. 

On the other hand, we measure nonzero values of $\xi$ between $2$ and $3~\mathrm{mHz}$ (see figure~\ref{fig:result_galaxy_demod}). This is the frequency range where the stochasticity of the overall stochastic \gls{gw} signal drops, due to the decreasing density of individual  \gls{ucbs} sources. This is consistent with other works that test the Gaussian and stationary assumptions in the same frequency range for the same kind of signal~\cite{riccardo,robbieTysonSGWB_stub}. However, given the above $\xi$ estimate (see Figure~\ref{fig:result_galaxy_demod}), the deviations of the residual data can be characterized with a distribution with only slightly wider tails compared to the Gaussian. Nevertheless, this slight deviation from Gaussianity could have an impact in parameter estimation and/or search for transient and monochromatic sources appearing at those frequencies. 

Finally, we should note again here that this result is dependent on the details of the analysis, i.e. the number of data segments $N_\mathrm{seg}$, the chosen whitening filter, or the binning of the spectrum. In addition, the confusion noise analyzed here is a product of a method that makes simplified assumptions about the resolvability of individual sources from the data. This means that the result of the \gls{snr}-based  iterative procedure of~\cite{PhysRevD.104.043019}, can be only seen as an output of an idealized parameter estimation procedure. Therefore, residual \gls{ucbs} data that have been generated with more realistic methodologies (e.g. from analyses that were presented in~\cite{Katz:2024oqg, Littenberg:2023xpl,Strub:2024kbe}), would yield a different result than the one presented in figures~\ref{fig:result_galaxy} and~\ref{fig:result_galaxy_demod}. 

\section{Discussion
\label{sec:discussion}}

We have adopted a flexible statistical framework for the characterization of stochastic signals based on the hyperbolic distribution, which was first introduced in the context of \gls{gw} data analysis in~\cite{Sasli:2023mxr}. One of the major advantages of this framework is that it can yield a robust estimate of the statistics of the residual data, which allows us to test for any possible departures from Gaussianity. This is done by computing the $\xi$ parameter (see eq.~(\ref{eq:xi})) as a combination of the parameters of the hyperbolic distribution~\cite{Eberlein2002TheGH,ernst}. We then apply this framework based on the hyperbolic likelihood, in order to fully characterize stochastic \gls{gw} signals for the future \gls{lisa} mission data. For our purposes, we have generated synthetic data sets of Gaussian cosmological signals, and non-stationary and non-Gaussian signals generated by the ensemble of \gls{ucbs} sources. 

In the first case, we demonstrate the methodology in fully Gaussian data for both the instrumental noise, and the signal from First-Order Phase Transitions. We fit the assumed unknown \gls{psd} of the data using shape-agnostic spline interpolation models, while at the same time estimating the hyper-parameters of the hyperbolic likelihood for testing for any departures from Gaussianity. As expected, the hyperbolic likelihood converges to the Gaussian likelihood, which yields the same posteriors on the parameters of the stochastic signal. 

In the second application, we focused on the stochastic signal generated by the many overlapping \gls{gw} signatures of \gls{ucbs}. This signal is expected to be cyclo-stationary, due to the orbital motion of the \gls{lisa} spacecraft constellation. We perform our analysis on the complete 4-year duration measurement, and we are able to detect this non-stationarity as a deviation from the Gaussian distribution: The $\xi$ parameter deviates from zero, especially in the frequency range where the stochastic signal is prevalent. In order to disentangle between heavy-tailed residuals due to nonstationarity and non-Gaussianity, we repeat the analysis by splitting the data into short time segments. Each week-long segment is then treated as independent, and its \gls{psd} is computed. We then assumed a joint hyperbolic likelihood for all segments, fitting common parameters as deviations from a spectral model for the general \gls{psd} of the data, as well as possible deviations from the Gaussian distribution. The results show that the non-stationarity of the signal is suppressed, but the $\xi$ parameter is deviating from zero between $2$ and $3~\mathrm{mHz}$. This indicates that the signal remains non-Gaussian in this area of the spectrum, something that was also verified in other studies~\cite{riccardo,robbieTysonSGWB_stub}. The non-Gaussianity in this region originates from the loss of stochasticity of the signal by the unresolved \gls{ucbs} due to their lower density and template overlap in each frequency bin. Almost all binaries above these frequencies are classified as resolved in this simulation~\cite{PhysRevD.104.043019}.

The above results demonstrate the practicality of the flexible statistical framework introduced in this work. In the same context, the hyperbolic likelihood can be employed to test the statistical properties of any given stochastic component of the data. Besides the characterization of stochastic signals, there are many other possible applications, such as the treatment of data irregularities. A prime example is the potential noise transients (glitches), especially when their rate is too high, and possibly their brightness too low to fit and remove each glitch individually from the data. Another possible application is the treatment of data gaps, where the flexible heavier-tailed hyperbolic distribution can be used to accommodate the effects of correlation between frequency bins due to the application of spectral windows on the gaps.

\appendix
\section{Supplementary Material}
\label{sec:appendix}
\begin{figure}[!htb]
    \centering
    \includegraphics[width=.9\textwidth]{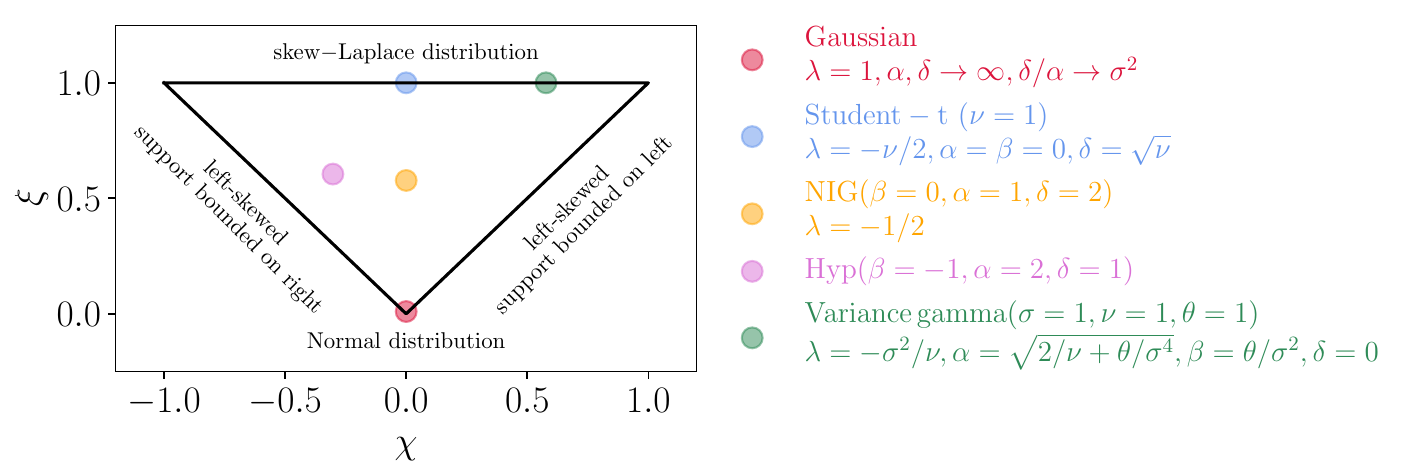}
    \caption{Example of a shape triangle plot of the $\chi-\xi$ parameter space for different distributions, with $\chi=\xi\beta/\alpha$. The shape triangle is a visualization tool that allows us to easily categorize a given distribution based on its overall shape. Each point represents a given distribution as a limiting distribution of the Generalized Hyperbolic one. In the legend, the Generalized Hyperbolic parameters are given in relation to the parameters of the distribution of each case. }
    \label{fig:toys}
\end{figure}

\acknowledgments

We would like to thank R. Rosati and M. Pieroni for the valuable discussions on the topic. N.K. acknowledges the funding from the European Union’s Horizon 2020
research and innovation programme under the Marie Skłodowska-Curie grant agreement No 101065596. R.B. acknowledges support through the Italian Space Agency grant \emph{Phase A activity for LISA mission, Agreement n. 2017--29--H.0}, by the MUR Grant ``Progetto Dipartimenti di Eccellenza 2023-2027'' (BiCoQ), and by the ICSC National Research Center funded by NextGenerationEU. A.S. acknowledges the Bodossaki Foundation for support in the form of a Ph.D. scholarship.\@

\paragraph{Software} The software used in the analysis can be found in \href{https://example.com}{github repository to be created}. The code for the iterative procedure of~\cite{PhysRevD.104.043019}, which was used in order to estimate the foreground confusion signal, can be found in \href{https://gitlab.in2p3.fr/Nikos/gwg}{gitlab.in2p3.fr/Nikos/gwg}. The \gls{mcmc} software used are the \href{https://github.com/mikekatz04/Eryn}{\textsc{Eryn}} sampler and a simple implementation of the Metropolis-Hastings algorithm which can be found in \href{https://gitlab.in2p3.fr/Nikos/metropolishastings}{gitlab.in2p3.fr/Nikos/metropolishastings}.

\bibliographystyle{JHEP}

\providecommand{\href}[2]{#2}\begingroup\raggedright\endgroup

\end{document}